\documentclass[conference]{IEEEtran}
\usepackage{cite}
\usepackage{amsmath,amssymb,amsfonts}
\usepackage{algorithmic}
\usepackage{graphicx}
\usepackage{textcomp}
\usepackage{xcolor}
\usepackage{siunitx}
\usepackage{url}
\usepackage{enumitem}
\usepackage{float}
\usepackage{placeins}
\usepackage{subcaption}
\usepackage{booktabs}
\usepackage{multirow}
\usepackage{array}
\newcolumntype{C}[1]{>{\centering\arraybackslash}p{#1}}
\usepackage{caption}
\usepackage{tikz}
\usepackage{pgfplots}
\usepackage{mathptmx}
\usepackage[protrusion=true,expansion=true]{microtype}
\usepackage{amsmath}
\pgfplotsset{compat=1.17}
\makeatletter
\g@addto@macro\normalsize{%
  \setlength\abovedisplayskip{3pt plus 1pt minus 1pt}
  \setlength\belowdisplayskip{3pt plus 1pt minus 1pt}
  \setlength\abovedisplayshortskip{0pt plus 1pt}
  \setlength\belowdisplayshortskip{2pt plus 1pt}}
\makeatother
\IEEEoverridecommandlockouts
\usetikzlibrary{arrows.meta,positioning,shapes.geometric}

\begin{document}

\title{Dynamically Reconfigurable MIMO-OFDM-LFM RFSoC Transceiver for Multifunctional Radars}

\author{\IEEEauthorblockN{Mradul Maurya, 
		Nanditha Unnikrishnan,
		Ankita V. Jaiswal,
		Ramkumar Raghu,
		R Rajesh}\\
		\IEEEauthorblockA{Center for Airborne Systems - DRDO, Bangalore, India\\\\
		$(mradul.maurya,nanditha.cabs,ankita.jaiswal06,ramkumarr.cabs,rajesh.cabs)@gov.in$}
}
\maketitle
\vspace{-14ex}  

\begin{abstract}
	Multiple-input multiple-output orthogonal frequency division multiplexing (MIMO-OFDM) is shown in literature to provide waveform agility necessary for multifunctional radar modes, including wide-area surveillance, imaging etc. Modulating OFDM subcarriers with Linear Frequency Modulation (LFM) improves range sidelobes in high-resolution range profiles (HRRP) and provides other benefits. We present an RFSoC-based MIMO-OFDM-LFM transceiver demonstrating seamless runtime radar mode switching in under 0.3 ms. We present an on-bench calibration method that achieves $0.25^\circ$ RMSE phase coherence across channels and subcarriers. We characterize beamforming, HRRP and beam-to-beam isolation performances. All measurements are done using standard Multichannel Vector Network Analyzer and Radar Target Simulator.
\end{abstract}


\section{Introduction}

Airborne multifunctional radars must support wide-area surveillance, high-resolution imaging, target tracking, and moving target indication (MTI) from a single aperture under tight size, weight, and power (SWaP) constraints. Conventional Active Electronically Scanned Arrays (AESAs), multifunctional operation necessitates complex radar resource management (RRM) and beam scheduling, as each radar function competes for the shared time resource. Fully digital arrays as in \cite{stralka2010,stralka2008wideband} use orthogonal frequency division multiplexing (OFDM) for simultaneous scheduling of multiple radar modes through independent digital beamforming on each subcarrier, supporting digital simultaneous looks (wide-area surveillance) in azimuth as well as concurrent MTI, and imaging from a single array. Recent theoretical studies, including circulating-code (CC) OFDM MIMO \cite{kuswidiastuti2022transmitter}, further explore subcarrier-based simultaneous azimuth scanning for digital radars. Such digital simultaneous looks have been shown to improve key RRM tasks, such as optimizing high-priority track scheduling during wide-area surveillance while freeing time resource for other modes of the radar \cite{unnikrishnan22}. Other advantages of using MIMO systems in radars, such as improved parameter identifiability, angular resolution, and transmit beampattern flexibility are studied in \cite{li07,fortunati20}.

Further, modulating each OFDM subcarrier with a Linear Frequency Modulated (LFM) baseband, OFDM-LFM brings added advantages. OFDM-LFM provides High-Resolution Range Profiles (HRRP) at a reduced per-subcarrier sampling rate \cite{liu22}. Optimal subcarrier phase and chirp-rate selections in OFDM-LFM suppresses periodic range sidelobes \cite{xia24,gi23}. Other advantages of OFDM-LFM such as clutter suppression through subcarrier amplitude weighting in colocated MIMO radar is studied in \cite{ding2022ofdm,ding2026signal}. Thus, MIMO OFDM-LFM is a strong candidate for Multifunctional Radars.

The practical implementation in the sense of multifunctional operation as in \cite{stralka2010,stralka2008wideband,noh2022,temiz2020} are limited. Transmitters for circulating-code (CC) OFDM MIMO radar \cite{kuswidiastuti2022transmitter} and FPGA-based OFDM-MIMO beamforming \cite{kumar24} have also been reported. OFDM-based high-resolution SAR-imaging was implemented by Wang \textit{et al.}\cite{wang2019first}. However. this is a sub-array architecture, unlike \cite{stralka2010,stralka2008wideband,kuswidiastuti2022transmitter}.

A full transceiver with low-sidelobe OFDM-LFM needs wideband converters on every channel, tightly coupled to logic that applies per-subcarrier coefficients at runtime. RFSoC provides this on one die. It has been used for true-time-delay beamforming \cite{steiner23}, full-array signal processing \cite{xiao23}, and synchronized digital-array signal generation and acquisition \cite{tian2025design}. Every channel–subcarrier pair must also be calibrated before coherent beamforming, and fast multicarrier transmit-array calibration is reported in \cite{kihira13}.

Existing designs are thus transmit-only \cite{kuswidiastuti2022transmitter,kumar24}, use fixed-waveform RFSoC beamformers \cite{steiner23,xiao23,tian2025design}, or assume ideal hardware \cite{xia24,ding2022ofdm,ding2023joint}. To the best of our knowledge, no work has fully implemented a MIMO-OFDM radar transceiver with $M$ transmit and receive channels on a single platform. In particular, none combines a chirp-capable OFDM baseband, runtime-reloadable subcarrier-level steering coefficients and per-channel, per-subcarrier calibration yielding phase-coherent waveforms.

This paper extends our work in~\cite{kumar24} to complete MIMO-OFDM-LFM RFSoC based transceiver with practical measurements, characterization, dynamic reconfigurability and bandwidth control for each subcarrier. We demonstrate runtime reconfigurability of the transceiver for wide-area surveillance and high-resolution imaging operations through the dynamic control. We also present an on-bench calibration method for the MIMO-OFDM transceiver with a multi-channel VNA, demonstrating subcarrier coherence across channels. Practical characterization of MIMO-OFDM-LFM beamforming and the beam-beam isolation and HRRP required for wide-area surveillance and imaging modes are also brought out.

\section{System Architecture}
\label{sec:sys_architecture}

\begin{figure}[H]
	\centering
	\includegraphics[width=0.8\columnwidth]{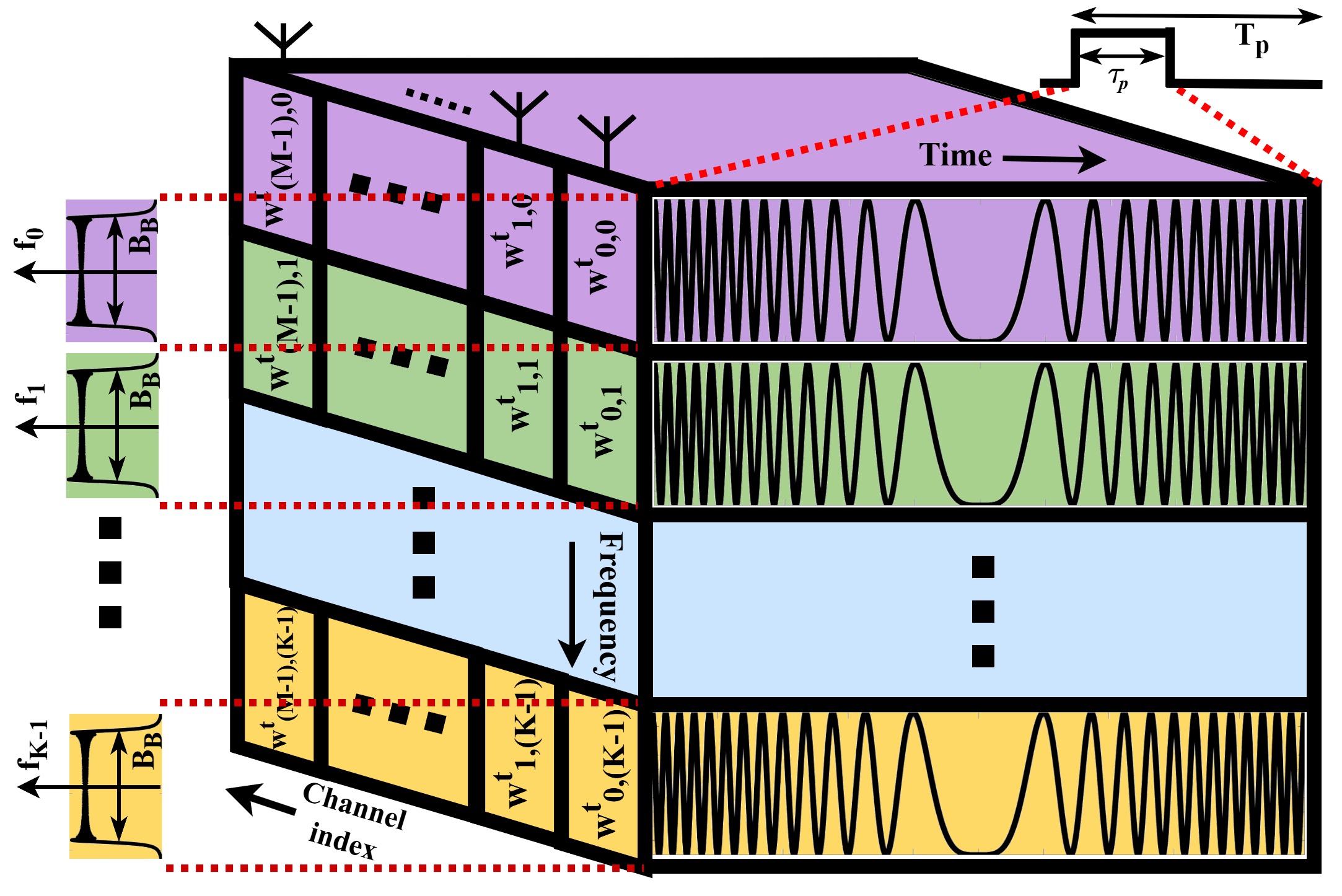}
	\caption{MIMO-OFDM subcarrier spatial allocation with LFM in transmit (ON-time $\tau_p$). Similar independent weighting applies to the receive chain of each subcarrier.}
	\label{fig:data_matrix}
\end{figure}

\begin{figure*}[t]
	\centering
	\includegraphics[width=0.85\textwidth]{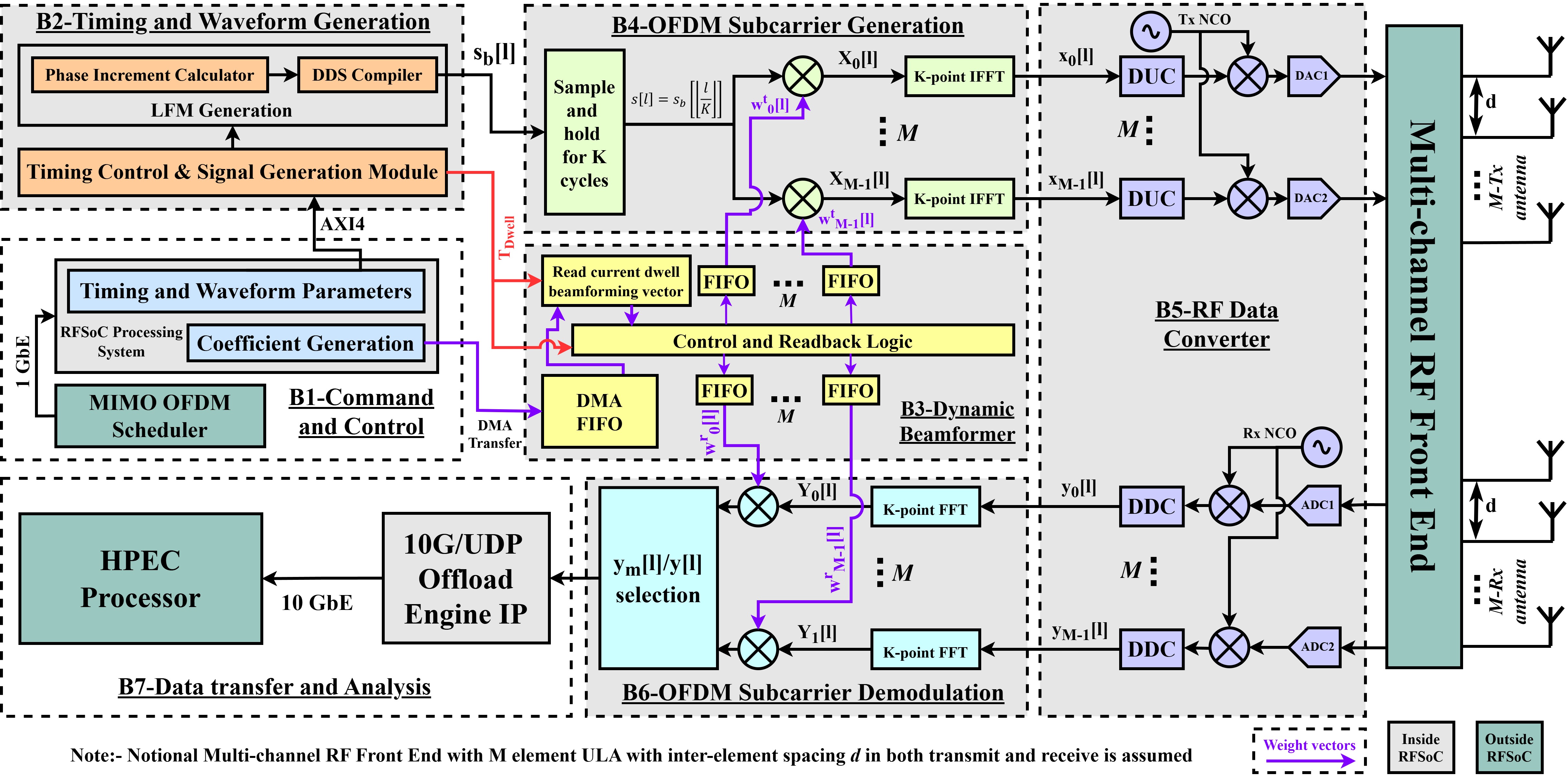}
	\caption{System block design of the proposed MIMO OFDM-LFM RFSoC transceiver}
	\label{fig:block}
\end{figure*}

We consider a MIMO-OFDM system with $K$ subcarriers, $M$-element Uniform Linear Array (ULA) with inter-element distance $d$ in both transmit and receive as shown in Figure \ref{fig:block}. A baseband LFM chirp waveform modulates all sub-carriers across all channels as shown in Figure \ref{fig:data_matrix}. The configurable chirp waveform is generated using Direct Digital Synthesizer (DDS). The baseband LFM is given as: 

\begin{equation}
s_{B}[n] = A e^{j \pi \mu \left(\frac{n}{f_{\text{clk}}}-\frac{\tau_p}{2} \right)^2}, \qquad 0 \le n \le \tau_p f_{clk}
\label{eq:lfm}
\end{equation}
where $A$ is the amplitude, $\mu$ is the chirp rate defined by LFM bandwidth $B_B$ and pulse width $\tau_p$ (as in $B_B = \mu \tau_p$) and $f_{clk}$ is the baseband sampling clock. The radar pulse repetition time (PRT) is $T_p$. All the parameters are controlled by the processor section of RFSoC. To modulate each subcarrier with this LFM pulse, every $K^{th}$ sample of $s_B[n]$ is latched for $K$ clock cycles as $s[l] = s_B[\lvert \frac{l}{K} \rvert]$. This limits LFM bandwidth to $f_{clk}/K$, which is also the maximum configurable limit of $B_B$. 

A Inverse Discrete Fourier Transform (IDFT) (implemented through AMD Xilinx IFFT IP) on the $K$ latched samples in $s[l]$ directly modulates the LFM $s[l]$ on all subcarriers. However, to introduce adaptive beamforming for each subcarrier, $s[l]$ is weighted with complex subcarrier beamforming weights, $w^t_{m,k} \in \mathbb{C}$, corresponding to $m^{th}$ channel and $k^{th}$ subcarrier, with $m \in [M], k \in [K]$ (With slight abuse of notation, $[N]$ means the set $\{0,1,\dots,N-1\}$ for any integer $N$) as follows:

Define stream of weights for $m^{th}$ channel as $w^t_{m}[l] = w^t_{m,k}$ when $l$ mod $K$ = $k$. This is realized by FIFO control (see Section \ref{sec:impl}). Now, $s[l]$ stream is multiplied by $w^t_m[l]$ as $X_m[l]=s[l] w^t_m[l]$. It is important to note that $s[l],w_m[l],X_m[l]$ streams consist of contiguous $K$ length frames. The $X_m[l]$ stream is directly fed to the AMD Xilinx IFFT block to generate the OFDM signal stream $x_m[l]$ as:

\begin{equation}
	x_m[l] = \frac{1}{K}\sum_{k=0}^{K-1} X^k_m[pl] e^{j 2\pi k  \frac{l}{K}},
	\label{eq:ifft}
\end{equation}

for every $p^{th}$ $K$ length frame of $X_m[l]$, $p=0,1, \dots,\lceil \frac{\tau_p f_{clk}}{K} \rceil$

The $\{ x_m[l], m \in [M] \}$ forms the transmitted MIMO OFDM signal shown in Figure \ref{fig:data_matrix}. The OFDM LFM is sequentially generated and gated to $[0,\tau_p]$ (within pulse-width duration) and repeated every $T_p$. The generated OFDM signal has subcarrier spacing $\Delta f = \frac{B}{K}$ where $B = f_{clk}$ is the total bandwidth of the OFDM signal. Further, the subcarriers are located at $f_k = k\Delta f$ for $k \in \{ 0,1,\dots,\frac{K}{2} \}$ and $f_k = (k-K)\Delta f$ for $k \in \{ \frac{K}{2},\frac{K}{2}+1,\dots,K \}$. Note that $K$ is always chosen to be a power of 2 due to the use of Radix-2 implementation in AMD Xilinx IFFT and FFT blocks. The baseband OFDM signal is upconverted to frequency $f_c$ using the internal NCO of RFSoC.

In practice, the coefficient $w^t_{m,k}$ combines the ideal steering coefficient $w^t_{\mathrm{ideal}}(m,k) \in \mathbb{C}$, calculated dynamically on the RFSoC processor by the MIMO-OFDM scheduler, and the channel phase and gain calibration coefficient $c^t_{\mathrm{cal}}(m,k) \in \mathbb{C}$ obtained through one time calibration (see Section \ref{sec:impl}):

\begin{equation}
	w^t_{m,k} = w^t_{\mathrm{ideal}}(m,k) c^t_{\mathrm{cal}}(m,k),
	\label{eq:calmul}
\end{equation}
\begin{equation}
	w^t_{\mathrm{ideal}}(m,k) = a_{m,k} e^{j\phi_{m,k}},
	\label{eq:steer}
\end{equation}

where $\phi_{m,k} = \frac{2\pi d\, m \sin\theta_k}{\lambda_k}, \forall m \in[M], k \in [K]$ with $\lambda_k=\frac{c}{f_c+f_k}$ and $d$ is the interelement distance of the ULA and $\theta_k$ is the steering angle for the $k^{th}$ subcarrier. The weights $w_{m,k}$ are computed in the processing section of RFSoC as per the steering requirement $\theta_k$, for each subcarrier. For eg., all subcarriers can be steered to the same angle ($\theta_k = \theta$) and all subcarriers can be integrated in receive (with equivalent receive subcarrier weighting) by simple DFT processing to obtain High Resolution Range Profile for imaging with resolution $\Delta r = \frac{c}{2B}$. Alternatively each $\theta_k$ can be distinct to obtain persistant wide area surveillance and each subcarrier data in receive can be processed concurrently. The range resolution in this case is $\Delta r = \frac{c}{2B_B}$. Any intermediate configuration is also feasible with subgrouping of subcarriers to different directions obtaining a range resolution of $\Delta r = \frac{c}{2n\Delta f}$ for $n$ contiguous subcarrier group. The amplitude term $a_{m,k}$ is added here to control the subcarrier amplitudes or mask subcarriers ($a_{m,k}=0$). 

Similarly, on the receive side, ADC stream $y_m[l]$ from the $m^{th}$ channel is passed through $K$-point FFT IP (AMD Xilinx FFT) for $m \in [M]$. The output $Y_m[l]$ of the FFT IP is weighted with independent weight stream $w^r_m[l]=w^r_{m,k}$ when $l$ mod $K=k$ for each $k \in [K], m \in [M]$. The weighted output stream is given by:

\begin{equation}
	y_m[l] = w_m[l]. Y_m[l] = w_m[l] \left( \sum_{k=0}^{K-1} y_m[pk]e^{-j2\pi k \frac{l}{K}} \right)
	\label{eq:fft}
\end{equation}

for every $p^{th}$, $K$-length frame of the input stream $y_m[l]$.

Now, it is essential to note that independent channel and subcarrier data is now available which can be either combined into a common stream $y[l] = \sum_{m=0}^{M-1} y_m[l]$ or can be independently streamed to the processor over multiple 10G lines discussed in Section \ref{sec:impl}. Though the latter is a multichannel stream useful for multichannel processing like STAP/MTI, it requires appropriate range gating to obtain a reasonable throughput/line rate for transfer over 10G lines. For eg., for $f_{clk}$ = 312.5 MHz, M=16, the throughput requirements of streaming all independent 32-bit streams is 160 Gbps (32*16*312.5), and for combined stream is 10 Gbps which can be comfortably transferred over two 10G lines. Alternatively, selective sub-array combining for each subcarrier is another option for controlling throughput albeit at the cost of reduced degrees of freedom.

It is also brought out that, the weights $w^t_{m,k}$ and $w^r_{m,k}$ can be configured for adaptive MIMO configurations including subarray MIMO \cite{li07}, phased MIMO \cite{ding2023joint} and massive MIMO \cite{fortunati20,bjornson19} configurations.

\section{Implementation}
\label{sec:impl}

We now bring out the implementation details of the MIMO-OFDM-LFM transceiver of Figure~\ref{fig:block}. The design is partitioned between the PS and PL of the RFSoC, with all PL blocks at $f_{clk}=312.5$ MHz. The MIMO-OFDM Scheduler on a Single Board Computer (SBC) (\textit{B1}) sends the timing ($\tau_p$, $T_p$), waveform ($B_B$) and beamforming parameters to the PS over 1 GbE. The PS computes the weights $w^t_{m,k}$, $w^r_{m,k}$ from \eqref{eq:calmul},\eqref{eq:steer}, combining steering with the stored/received gain-phase correction vector from one-time calibration, where $c^t_{\mathrm{cal}}[m,k]$ is the inverse of the response measured per channel and subcarrier by a multichannel VNA. Waveform and timing parameters go to the PL over a custom AXI4 control interface (\textit{B2}), where a timing control module generates the PRT and $T_{dwell}$ signals and triggers an RTL block that generates the accumulator phase for the AMD Xilinx DDS Compiler IP, which outputs the complex LFM baseband $s_B[n]$ of \eqref{eq:lfm} (bandwidth $B_B$, duration $\tau_p$) as an AXI4-Stream, one sample per clock.

All PL streams are AXI4-Stream, with one OFDM symbol framed as $K$ samples (see section \ref{sec:sys_architecture}) and TLAST on every last $K^{th}$ sample. In \textit{B4}, each sample of $s_B[n]$ is held for $K$ clocks to form $s[l]$. Weights reach the PL via AXI DMA into a DMA FIFO (\textit{B3}), enabling high-speed mode switching and keeping steering angle and waveform timing configuration independent. At the rising edge of $T_{dwell}$, the dwell vectors are read into each channel's FIFO under synchronised read enables from a custom Control and Readback logic, which also loads new coefficients or reads back the loaded ones; a new configuration is always applied at the next PRT edge. Each FIFO is read cyclically every $K$ samples to stream $w^t_m[l]=w^t_{m,k}$, $l \bmod K=k$, framed like $s[l]$ and issued against the same frame counter, so the complex multiplier receives both inputs on a joint TVALID, pairing $l^{th}$ sample with corresponding $l^{th}$ weight and outputs $X_m[l]=s[l]w^t_m[l]$ with TVALID and TLAST delayed by the multiplier's fixed latency. The AMD Xilinx FFT IP, configured as IFFT, runs in pipelined streaming mode, taking and returning one sample of $x_m[l]$ per clock with a fixed latency and without loss of throughput. $x_m[l]$ goes to the synchronised DACs via the RF Data Converter IP (\textit{B5}), whose DUC, DDC and NCO are configurable, the NCO upconverting the waveform to $f_c$.

\begin{figure*}[t]
	\centering
	\includegraphics[width=\textwidth]{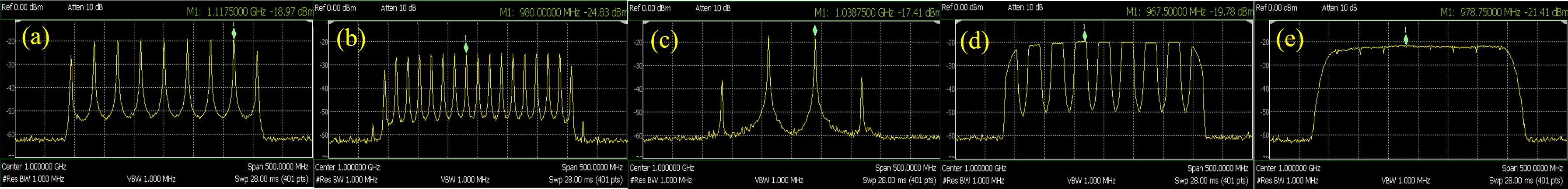}
	\caption{Dynamic configurability of OFDM parameters : (a) 8 subcarriers with 0 Hz baseband, (b) 16 subcarriers with 0 Hz baseband, (c) masking of subcarriers with amplitude variation, (d) 8 subcarriers with LFM baseband of $B_B/2$ subcarrier bandwidth, (e) 8 subcarriers with LFM baseband of $B_B$ subcarrier bandwidth. (Note: These OFDM spectra are without FFT-shift after IFFT (Figure \ref{fig:block}). Also the spectral envelope as seen in (e) is due to the response of filters in DDC chain of RFSoC)}
	\label{fig:configurability}
\end{figure*}
	
\begin{figure*}[t]
	\centering
	\begin{subfigure}[t]{0.39\textwidth}
	\centering
	\includegraphics[width=\linewidth]{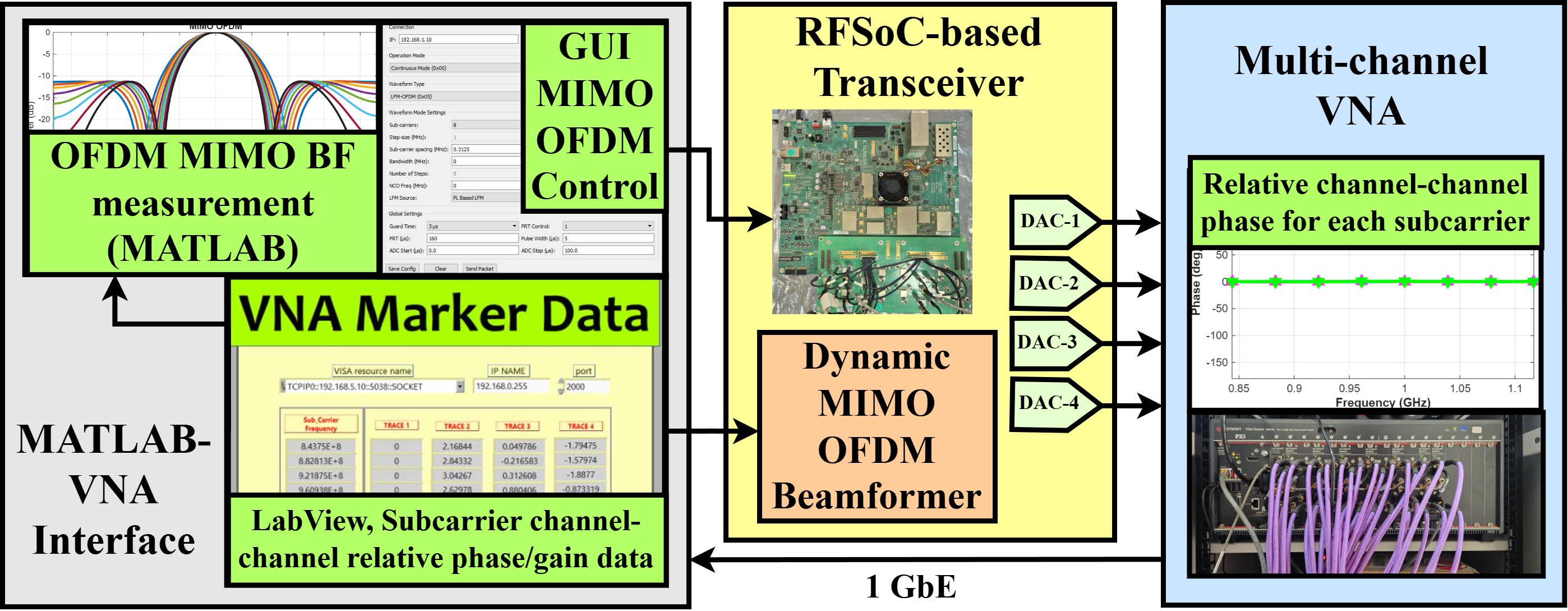}
	\caption{}
	\label{fig:exp_setup1}
	\end{subfigure}
	\hfill
	\begin{subfigure}[t]{0.59\textwidth}
	\centering
	\includegraphics[width=\linewidth]{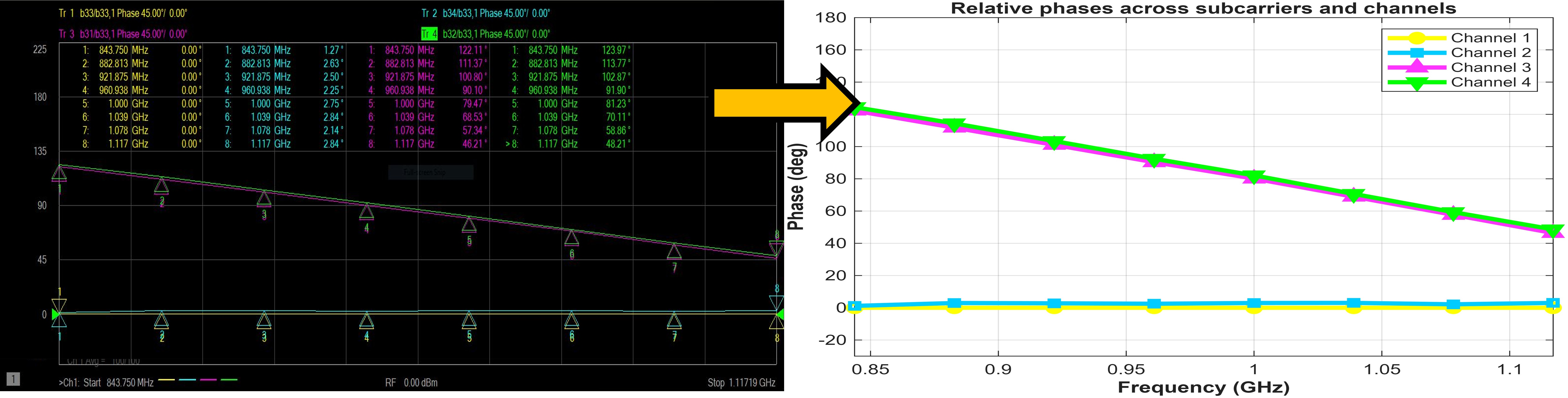}
	\caption{}
	\label{fig:uncomp}
	\end{subfigure}
	\caption{(a) Experimental block diagram for MIMO OFDM-LFM beamforming using measured phases and array factor; (b) uncompensated state of the system (VNA marker data and relative phases plotted in MATLAB) when absolute zero phase is applied to all channels and subcarriers.}
	\label{fig:cal_setup}
\end{figure*}

On receive, the ADC stream $y_m[l]$ from \textit{B5}, after the DDC (DAC to ADC loopback path has fixed latency), is framed into $K$-sample frames by the timing logic. In \textit{B6}, the channel-wise AMD Xilinx FFT IP, computes the $K$-point FFT, giving $Y_m[l]$ with TLAST on each $p^{th}$ output frame's last sample. The weight stream $w^r_m[l]$ is read from the \textit{B3} FIFOs in the same cyclic manner, aligned to the dwell and PRT timing, and the receive complex multiplier forms $w^r_m[l]Y_m[l]$ of \eqref{eq:fft}, i.e. receive digital beamforming per subcarrier. Summing over $M$ channels gives $y[l]$, and the $y_m[l]/y[l]$-selection block enables choice of range-gated multi-channel data or beamformed sub-carrier data for 10GbE link.

The selected stream goes over 10GbE to the external HPEC processor through the 10G/UDP Offload Engine IP \cite{ankita2026} (\textit{B7}) for multi-channel sub-carrier receive calibration (yielding $c^r_{\mathrm{cal}}(m,k)$), pulse compression, HRRP characterization and RTES-based receive beamforming characterization.

\textit{CP, PAPR, and ICI considerations~:~} For $B_B=0$, repeated OFDM symbols across subcarriers act as cyclic prefixes; for $0<B_B\le \Delta f$, echoes remain in their own FFT bins and range is resolved by pulse compression\cite{knill2021}, avoiding a dedicated cyclic prefix in our implementation. PAPR is reduced by an offline-optimized per-subcarrier phase $\theta_k$ \cite{liu2024} loaded via the existing DMA path to the \textit{B3} . ICI is managed by selecting $B_B<\Delta f$ and disabling alternate subcarriers via $a_{m,k}$ \cite{yazgan2024}.

\section{Results and discussion}
\label{results}

This section brings out the results of the experiments performed to validate the proposed RFSoC-based MIMO OFDM-LFM transceiver implemented on ZCU216 RFSoC board, and configured for 16 channels. But due to balun limitations, only 4 channel MIMO-OFDM is demonstrated. For conducting the experiments, the sampling rate $(f_s)$ is 2.5 GSPS, NCO frequency $(f_c)$ is 1 GHz, $(f_{clk})$ and total bandwidth $(B)$ is 312.5 MHz, number of subcarriers $(K)$ is 8 (16 in RTES test), subcarrier bandwidth $(B_B)$ is $\frac{B}{K}$, pulse width $(\tau_p)$ is 5~$\mu$s, and pulse repetition time $(T_p)$ is 1~ms.

\subsection{Dynamic subcarrier and bandwidth control}

We demonstrate the dynamic configurability of the proposed system in terms of number of subcarriers, baseband waveform and masking of subcarriers. All waveform spectra in Figure~\ref{fig:configurability} are taken from Keysight N9917A Spectrum Analyzer connected to a single channel DAC output of the RFSoC transceiver. Note that due to lack of FFT shift operation in IFFT IP, the $\left( \frac{K}{2}+1 \right) ^{th}$ subcarrier lies at the band edge, its energy is split into 2 peaks at $f_k=-B/2$ and $f_k=B/2$. This is due to the filter response of the DUC/DFE chain of the RFSoC. Further, Figure~\ref{fig:configurability}(c) demonstrates subcarrier amplitude controls where the alternate subcarriers are disabled by making $a_{m,k}=0$ for $k=odd$. For $k=even$, amplitude taper is applied. Such kind of amplitude controls are useful for spectral shaping as in \cite{ding2022ofdm,ding2023joint} to suppress clutter and optimize the OFDM spectrum for target. Figures~\ref{fig:configurability}(d) and (e) demonstrate adaptive bandwidth control for the subcarrier LFM signal with $B_B=19.53125~MHz$ and $B_B=39.0625~MHz$ respectively. The bandwidth control is essential for ICI control and we will see that $B_B$ directly controls the HRRP sidelobes in Section \ref{sec:highbwimag}.

\subsection{Multi-channel Sub-carrier calibration}

Figure \ref{fig:exp_setup1} shows the benchtop setup used to estimate the residual phase and gain errors and, subsequently, to characterize real-time beamforming. Four DAC channels are connected to a Keysight M9804 multichannel VNA. The MIMO-OFDM scheduler applies zero phase $(\phi_{m,k} = 0)$ and uniform gain $(a_{m,k} = 127)$ to all subcarriers, using a single-tone baseband $(B_B = 0)$. Since the signals are pulsed, the VNA uses a narrow IF bandwidth $(< 1~kHz)$ to reduce phase noise. It measures at $K$ spot frequencies, one per subcarrier.

Instead of conventional S-parameter measurement, the VNA's math function computes the receiver-trace ratio $\frac{b_{m,k}}{b_{1,k}}$. This ratio is the relative phase and gain of each channel-subcarrier pair with respect to the first channel at the same subcarrier. The calibration coefficients are its inverse: $c^t_{cal}(m,k) = \left(\frac{b_{m,k}}{b_{1,k}}\right)^{-1}$.

\begin{figure*}[t]
	\centering
	\includegraphics[width=\textwidth]{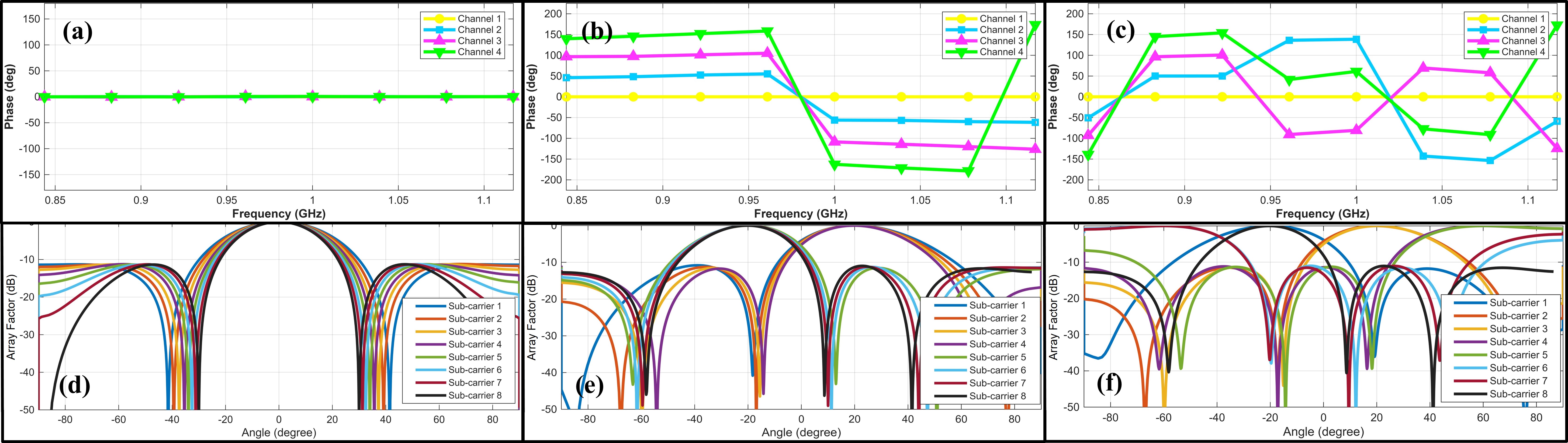}
	\caption{VNA-measured relative phases across channels and subcarriers (top row) and the corresponding array factor patterns using VNA-measured excitations (bottom row): all subcarriers looking at $0^\circ$ (a, d), subcarriers looking in two directions $-20^\circ,20^\circ$ (b, e), and subcarriers looking in four directions $-60^\circ,-20^\circ,20^\circ,60^\circ$ (c, f).}
	\label{fig:pattern_six_panels}
\end{figure*}

A custom LabVIEW application fetches these coefficients using VISA and SCPI protocol, and transfers to the processing system which stores them in memory so that recalibration is not needed at every power-on. Figure \ref{fig:uncomp} shows the uncompensated state. Figure \ref{fig:pattern_six_panels}(a) shows the compensated state, in which residual phase and gain errors are removed across all channels and subcarriers, making the subcarriers coherent across channels. The calibration table was further validated across power cycles, giving an RMSE of $0.25^\circ$ per subcarrier over 100 snapshots.

\subsection{Dynamic MIMO-OFDM-LFM Beamforming for Multifunctional Operation}

The subcarrier weights (phase/gain) can be dynamically controlled at runtime, enabling seamless switching to multiple radar modes. The MIMO-OFDM scheduler sends the look-directions for each sub-carrier and the final sub-carrier transmit/receive coefficients are calculated at the processing system which further gets multiplied with the stored calibration coefficients and then loaded to the dynamic Beamformer block B3 as explained in Sections \ref{sec:sys_architecture}, \ref{sec:impl}. Through the experiments, we have characterized the performance of two major applications of the multi-functional MIMO-OFDM transceiver.

\subsubsection{High-Bandwidth Imaging}
\label{sec:highbwimag}

All subcarriers, or a contiguous subset, can be steered to a common direction $(\theta_k = \theta)$. Their echoes are then coherently integrated in receive to synthesize a wideband waveform. Steering all $K$ subcarriers yields the full bandwidth $B$ = 312.5 MHz ($\Delta r = \frac{c}{2B} \approx$ 0.48 m). A group of $n$ contiguous subcarriers provides a bandwidth of $n \Delta f$ and a resolution of $\Delta r = \frac{c}{2n\Delta f}$. The design is validated in two steps: (i) Characterizing the spatial coherence required for aligning all sub-carriers in one direction: through real-time array factor plotted in MATLAB, (ii) Characterizing the synthesized high bandwidth required for achieving high resolution: through HRRP.

\textit{Spatial Coherence~:~} The setup of Figure \ref{fig:exp_setup1} is used. The relative phases measured by the multichannel VNA pass through a LabVIEW interface to a MATLAB script, which computes the array factor in real time. Figures \ref{fig:pattern_six_panels}(a) and (d) confirm that all eight subcarriers are steered to a common direction $0^\circ$. Figures \ref{fig:pattern_six_panels}(b) and (e) show two subcarrier groups steered to $\pm 20^\circ$, which enables imaging in two directions simultaneously, thereby resulting in coherently stitched bandwidth of 156.25 MHz with a range resolution of $\Delta r \approx 0.96 m$ simultaneously in both directions.

\begin{figure}[H]
	\centering
	\includegraphics[width=\columnwidth]{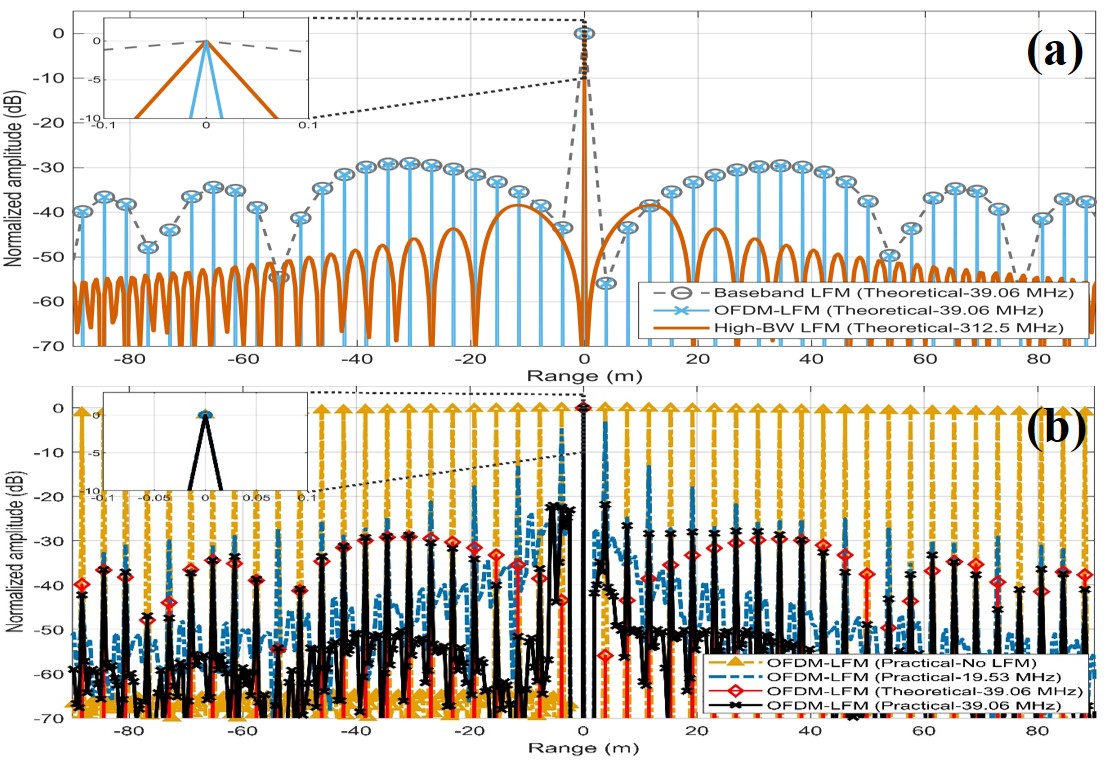}
	\caption{HRRP comparison of high-bandwidth LFM, baseband LFM and OFDM LFM with varying subcarrier bandwidth}
	\label{fig:hrrp}
\end{figure}

\textit{Synthesized Bandwidth~(HRRP)~:~} Now, we characterize the synthesized HRRP required for high resolution imaging mode. In this experiment, all subcarriers are phase-aligned, and the waveform is looped back from a single DAC to a single ADC channel, post which it is demodulated and streamed over 10 GbE to the HPEC processor for HRRP creation. Figure \ref{fig:hrrp}(a) compares the theoretical HRRP of OFDM-LFM with that of high-bandwidth LFM at 312.5 MHz and 39.0625 MHz. Figure \ref{fig:hrrp}(b) compares the measured HRRP of the 8-subcarrier waveform for $B_B$ = 0, $\frac{B}{16}$ and $\frac{B}{8}$.

\begin{figure*}[t]
	\centering
	\begin{subfigure}[t]{0.29\textwidth}
	\centering
	\includegraphics[width=1.1\linewidth]{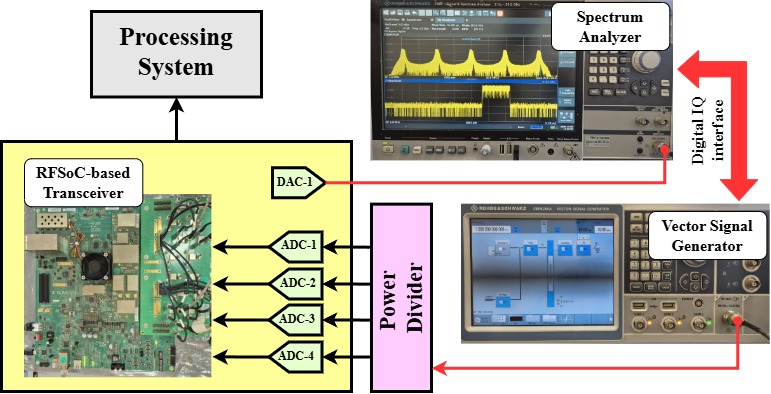}
	\caption{}
	\label{fig:rtes_setup}
	\end{subfigure}
	\hfill
	\begin{subfigure}[t]{0.34\textwidth}
	\centering
	\resizebox{1.1\linewidth}{!}{\input{rtes_receive_bf.tex}}
	\caption{}
	\label{fig:rtes_pattern}
	\end{subfigure}
	\hfill
	\begin{subfigure}[t]{0.34\textwidth}
	\centering
	\includegraphics[width=0.9\linewidth]{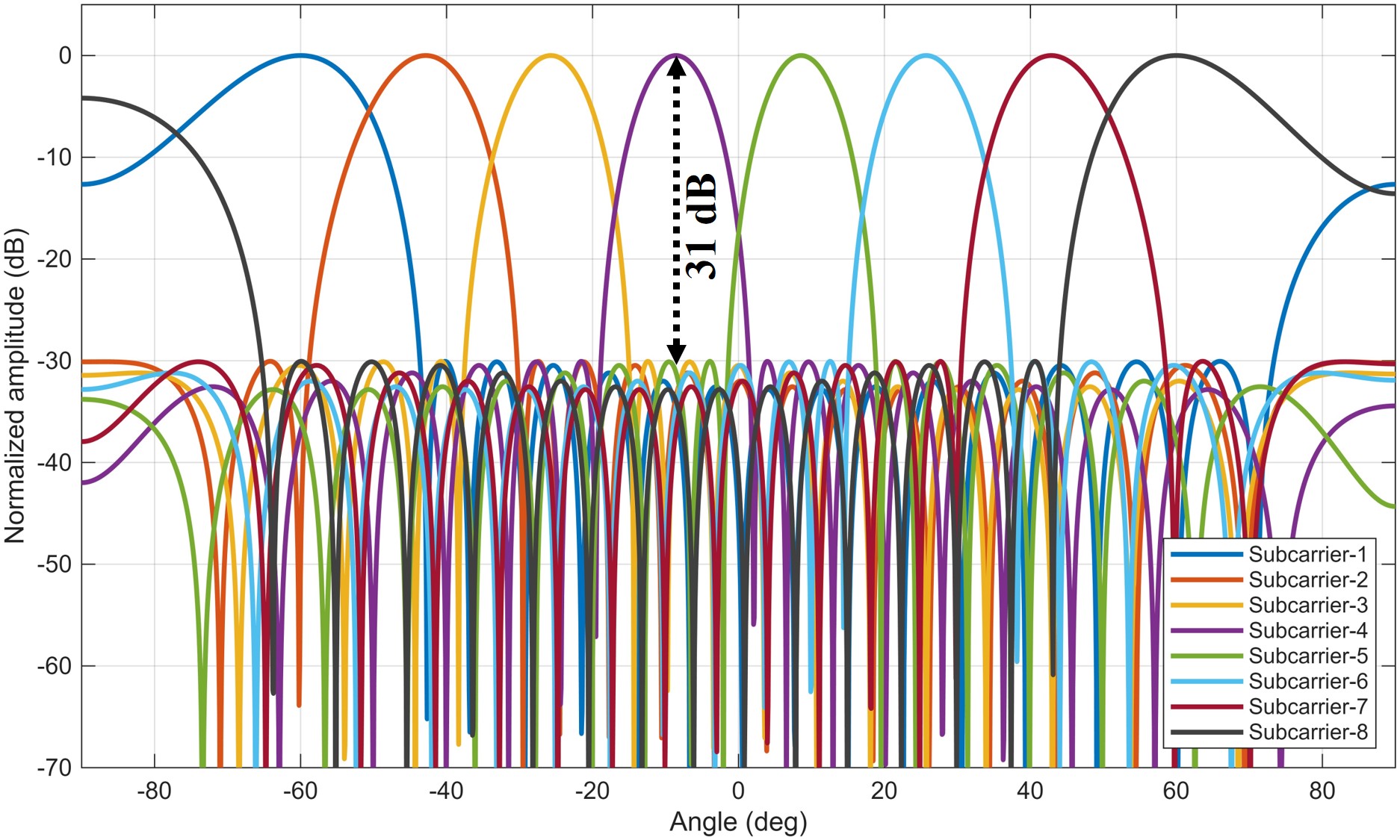}
	\caption{}
	\label{fig:16ch_pattern}
	\end{subfigure}
	\caption{(a) Experimental setup block diagram for receive beamforming characterization using RTES; (b) measured leakage power of other subcarriers with respect to Subcarrier 4, which is looking in boresight, using RTES; (c) simulated theoretical array factor for a 16-channel ULA with 8-subcarrier OFDM-LFM and Taylor's window applied (SLL $-35$ dB, $\bar{n}=4$).}
	\label{fig:rtes_all}
\end{figure*}

It can be observed from Figure \ref{fig:hrrp}(b) that baseband LFM bandwidth is the key to range-sidelobe control. For $B_B = 0$, the subcarriers form a discrete frequency comb. Its response contains periodic sidelobes, spaced at $\frac{c}{2\Delta f} \approx 3.84$ m, that reach main-lobe level. Chirping each subcarrier over $B_B$ fills the spectral gaps between adjacent subcarriers, which flattens the composite spectrum and suppresses these peaks. As $B_B$ increases from 0 to 19.53 $(\frac{B}{16})$ MHz and then 39.06 $(\frac{B}{8})$ MHz, the sidelobe metrics improve as follows:

\begin{itemize}
	\item PSLR improves from 0 to -3.10 and -21.8 dB.
	\item ISLR improves from 16.10 to 0.23 and -12.07 dB
	\item ASL improves from -9.57 to -25.45 and -37.77 dB
\end{itemize}

At $B_B = \Delta f$, the spectrum is contiguous. The theoretical response then matches that of a critically sampled LFM of bandwidth $B$ (PSLR $\approx$ -38.4 dB). The remaining deviation of the measured response is attributed to the fractional delay between the ADC sampling instant and the true echo peak. Sidelobe level is therefore directly controllable through $B_B$. As brought out through Figure \ref{fig:hrrp}, the OFDM-LFM adopts the good properties of LFM through lower sidelobes, while synthesizing the full bandwidth $B$ within a single PRT, without the time-on-target penalty associated with stitched waveforms such as stepped LFM.

\subsubsection{Simultaneous Multi-Directional Looks}

For persistent wide-area surveillance, each subcarrier or subcarrier group is assigned a distinct steering angle $\theta_k$ and processed concurrently. Each look has a range resolution of $\frac{c}{2B_B}$. The performance of this mode is also in done in two ways: (i) Characterizing the real-time array factor for different look angles (ii) Characterizing the beam-to-beam isolation achieved when sub-carriers look in non-overlapping directions.

\textit{Real-time array factor~:~} Figures \ref{fig:pattern_six_panels}(c) and (f) show subcarriers steered to $-60^\circ$, $-20^\circ$, $20^\circ$ and $60^\circ$, each forming a distinct main lobe. The directions of each sub-carrier is chosen such that there is no overlap in the 3 dB beamwidth of each beam. This mode aids in simultaneous surveillance of a wide area.

\textit{Beam-to-beam isolation~:~} Receive beam-to-beam isolation is characterized with the Radar Target Echo Simulator (RTES) setup of Figure \ref{fig:rtes_setup}: consisting of FSW26 spectrum analyser and SMW200A vector signal generator. A single DAC output drives the RTES, which emulates a target echo at $0^\circ$. The echo is distributed to four ADC channels through a 1:4 power divider and phase-matched cables. Of the $K = 16$ subcarriers, 8 are enabled because the RTES analysis bandwidth is limited to 200 MHz. These eight are steered to $-52.6^\circ$, $-52.6^\circ$, $-26.3^\circ$, $0^\circ$, $26.3^\circ$, $26.3^\circ$, $52.6^\circ$ and $52.6^\circ$. Figure \ref{fig:rtes_pattern} compares the matched-filtered output power of each subcarrier with its theoretical array factor. Subcarrier 4, steered to $0^\circ$, peaks at boresight. The boresight leakage of the remaining subcarriers follows their array factors, giving a minimum isolation of 12.4 dB.

Figure \ref{fig:16ch_pattern} shows that a 16-channel array with Taylor weighting (SLL -35 dB, $\bar{n}$ = 4) raises the isolation to about 31 dB, at the cost of a wider main lobe. Applying Golay or equivalent codes \cite{kuswidiastuti2022transmitter} to the phase/gain vectors can improve isolation further.

\subsection{Resource Utilization}

On the ZCU216 with 16 channels, the transceiver uses LUT $(29.28\%)$, LUTRAM $(10.53\%)$, FF $(20.56\%)$, BRAM $(6.11\%)$, DSP $(8.99\%)$, IO $(2.56\%)$, BUFG $(2.44\%)$, MMCM $(25\%)$. This RFSoC implementation addresses one of the possible upgradations mentioned in~\cite{kumar24}, and the moderate utilization relative to the available resources leaves headroom for other waveforms, signal processing algorithms etc. for other modes of operation. The Spurious Free Dynamic Range (SFDR) is 63 dB, measured on FSW26 Spectrum Analyzer (RBW 10 Hz, span 500 MHz) with OFDM configured for 8 subcarriers and a 0 Hz baseband signal. Runtime switching between surveillance, high-resolution imaging and split-direction operation, a key feature which also makes the transceiver configurable for MTI, takes $<0.3$ ms, the total time for coefficient generation, DMA-based PS-to-PL transfer and OFDM subcarrier modulation, obtained from PL counter values at 3.2 ns per clock cycle; new beamforming vectors are applied at the next PRT, avoiding any data loss and corruption. The architecture also caters for constant modulus baseband waveforms for low Peak-to-Average Power Ratio (PAPR), although this is not included in the scope of this work.

\section{Conclusion and Future Work}

This paper described a dynamically configurable MIMO OFDM-LFM RFSoC transceiver in which a baseband LFM waveform modulates orthogonal OFDM sub-carriers. It dynamically reconfigures PW, PRT, bandwidth and number of sub-carriers, and seamlessly switches between surveillance, high-resolution imaging and split-direction (simultaneous multi-direction) modes with $<0.3$ ms latency, as demonstrated. With appropriate optimal weighting of sub-carriers, enhanced clutter suppression capabilities are also achieved. Automated multi-channel, multi-subcarrier Tx/Rx calibration achieves RMSE $0.25^\circ$ per subcarrier. HRRP characterization showed significant PSLR and ISLR improvement with the incorporation of LFM in baseband, and RTES tests of multi-directional looks showed that around 31 dB beam-to-beam isolation is achievable with appropriate windowing of receive phase/gain weights. Planned extensions include antenna range-based far-field validation to complement the VNA measurements, operation across a larger fraction of the platform's 16 physical transmit and receive channels, Doppler/MTI validation using multipulse data, and validation on opportunistic airborne targets.

\section*{Acknowledgment}

The authors sincerely acknowledge Mrs. P. Santhya, Outstanding Scientist and Director-CABS, for her motivation and support, and for facilitating the communication of this paper to the IEEE RADAR-Conf-2027.

\bibliographystyle{IEEEtran}
\bibliography{bibliography}

\end{document}